\documentclass[twocolumn,aps,prb,floats,superscriptaddress,showpacs]{revtex4}
\usepackage{amsmath}
\usepackage{epsfig}
\usepackage{graphicx}
\usepackage{dcolumn}
\usepackage{bm}
\usepackage{color}

\def\gapp{\lower.35em\hbox{$\stackrel{\textstyle>}{\sim}$}}
\def\lapp{\lower.35em\hbox{$\stackrel{\textstyle<}{\sim}$}}

\begin{document}

\title{{Fluctuation-induced dispersion forces on thin DNA films}}
\author{Lixin Ge}
\email{lixinge@hotmail.com}
\affiliation{School of Physics and Electronic Engineering, Xinyang Normal University,
Xinyang 464000, China}
\author{Xi Shi}
\affiliation{Department of physics, Shanghai Normal University, Shanghai, 200234, China}
\author{Binzhong Li}
\affiliation{School of Physics and Electronic Engineering, Xinyang Normal University,
Xinyang 464000, China}
\author{Ke Gong}
\affiliation{School of Physics and Electronic Engineering, Xinyang Normal University,
Xinyang 464000, China}
\date{\today}

\begin{abstract}
In this work, the calculation of Casimir forces across thin DNA films is
carried out based on the Lifshitz theory. The variations of Casimir forces
due to the DNA thicknesses, volume fractions of containing water, covering
media and substrates are investigated. For a DNA film suspended in the air
or water, the Casimir force is attractive, and its magnitude increases
with decreasing the thickness of DNA films and the water
volume fraction. For DNA films deposited on a dielectric(silica) substrate,
the Casimir force is attractive for the air environment. However, the Casimir force shows unusual features in a water environment. Under specific conditions, switching signs of the Casimir force from attractive to repulsive can be
achieved by increasing the DNA-film thickness. Finally, the Casimir force for DNA films deposited on a metallic substrate are investigated. The Casimir force is dominant by the repulsive interactions at a small DNA-film thickness for both the air and water environment. In a water environment, the Casimir force turns out to be attractive at a large DNA-film thickness, and a stable Casimir
equilibrium can be found. In addition to the
adhesion stability, our finding could be applicable to the problems of
condensation and de-condensation of DNA, due to the fluctuation-induced
dispersion forces.
\end{abstract}

\maketitle


\section{Introduction}

The dispersion force is generated by the fluctuating dipoles, resulting from the zero-point vacuum fluctuation and thermal fluctuation \cite{Fei:89}. When the consuming time for propagating waves between the fluctuation dipoles is larger or comparable with the lifetime of fluctuating dipoles, the retardation effect (or wave effect) can modified the separation-dependence decaying laws of dispersion force \cite{Rod:11}. Specifically, the dispersion force is known as the van der Waals force for
closely spaced objects or interfaces \cite{Par:05}, where the retardation is negligible. The retardation effect manifests when the separation distance is large, and the dispersion force is also named as retarded van der Waals force\cite{Dag:02} or the Casimir force
\cite{Bor:09, Kli:09}. In some configurations, the retardation effect can be apparent even at the separation of several nanometers\cite{Lee:01}. The dispersion force and its free energy play an important role in various disciplines, ranging from nanomechanics\cite{Gon:21,Som:18,Ge:20b,Mun:21,Est:15}, wetting phenomena\cite{Hou:80,Boi:11,Squ:22}, to ice pre-melting and formation \cite{Li:22,Lue:21,Est:20,Fie:20} etc. In addition, the dispersion force and its free energy across organic films  were
also investigated intensely \cite{Lu:15, Bla:21, Bar:19,Kli:20,Kli:21,Kli:22}. It was
reported that the attractive dispersion force would make the organic films more
stable, while the repulsive force has an opposite effect\cite{Bar:19,Kli:20,Kli:21,Kli:22}.

Deoxyribonucleic acid (DNA) composed of two helical polynucleotide chains is one of the most important substances in biology. Along with its biological functions, the material properties of DNA are of great interest for the state-of-art of nanotechnology, motivated by the promising applications in a variety of fields, such as self-assembly of colloidal nanoparticles \cite{Nyk:08,Cho:14,Rog:16,Cui:22}, DNA-based nanomedicines\cite{Cam:10,Hu:18,Wei:21,Gu:21}, organic nanophotonics \cite{Ste:07, Bui:19} etc. As one of the crucial elements, the DNA films have been widely applied in many bio-organic nanoscale devices \cite{Kaw:00, Zha:12, Kha:17,Jun:17}. The DNA films are generally deposited on inorganic substrates using the spin coating process \cite{Jun:17}. The structure stability of double-stranded DNA is determined by the hydrogen bonds between nucleotides and the base-stacking interactions\cite{Yak:06}. However, the adhesion stability of DNA films placed on a substrate is dependent on the surface forces \cite{Boi:11}, such as ionic or electrostatic forces, intra-hydrogen bonds, dispersion forces etc. The dispersion force is an important ingredient at the surface forces, particularly, when the thickness of a bio-organic film is miniaturized to a sub-micro scale\cite{Bar:19}. Moreover, the other surface forces (e.g., intra-hydrogen bonds) could be absent at the surface of some specific substrates. Then, it is expected that the contribution from the dispersion force becomes more prominent, and the quantitative calculations of this force are necessary.

In this work, we study the Casimir force of DNA films within the framework of
Lifshitz theory. The influences of Casimir forces due to DNA-film
thicknesses, water volume fractions, background media and
substrates are investigated by numerical calculations. We find that the Casimir pressure is attractive for a DNA film suspended in the air or water, and its magnitude
increases by decreasing the DNA-film thickness and water volume fraction. For a DNA film placed on a silica substrate in the air background, the Casimir pressure shows a similar trend as that in the suspended configuration. However, the Casimir pressure exhibits rich features when the setup is immersed in the water. Under specific water volume fraction, switching sign
of the Casimir pressure across a wet DNA film is revealed by increasing the DNA-film thickness. Finally, the Casimir pressure for a DNA film placed on a metallic substrate is also calculated. It is found that the Casimir pressure is dominant by the repulsive interactions at a small DNA-film thickness for both the air and water environment. Interestingly, a stable Casimir equilibrium is found when the DNA film is immersed in the water. Our findings could be applicable to the problems of adhesive stability, condensation and de-condensation of DNA films, due to the fluctuation-induced
dispersion forces.

\section{Theoretical models}

We consider a DNA film with thickness $a$ sandwiched between a cladding
medium and a substrate. The thicknesses of the cladding layer and substrate
are assumed to be semi-infinite. In addition, the whole system is in thermal
equilibrium at room temperature $T$. The Casimir pressure of the DNA film is
calculated based on the framework of the Lifshitz theory \cite{Kli:09,
Kli:20}:
\begin{equation}
P_{\mathrm{c}}(d)=-\frac{k_{b}T}{\pi }\overset{\infty }{\underset{n=0}{\sum }%
}^{\prime }\int_{0}^{\infty }k_{\Vert }k_{3}dk_{\Vert }\underset{\alpha =s,p}%
{\sum }\frac{r_{1}^{\alpha }r_{2}^{\alpha }e^{-2k_{3}a}}{1-r_{1}^{\alpha
}r_{2}^{\alpha }e^{-2k_{3}a}},
\end{equation}%
where the prime in summation denotes a prefactor 1/2 for the term $n=0$, $%
k_{b}$ is the Boltzmann's constant, $k_{3}=\sqrt{k_{\parallel
}^{2}+\varepsilon_{\mathrm{D}}(i\xi _{n})\xi _{n}^{2}/c^{2}} $ is the
vertical wavevector in the DNA film, $k_{\parallel}$ is the parallel
wavevector, $c$ is the speed of light in vacuum, $\varepsilon _{\mathrm{D}%
}(i\xi _{n}) $ is the permittivity of the dry DNA film, $\xi _{n}=2\pi \frac{%
k_{b}T}{\hbar }n(n=0,1,2,3...)$ are the discrete Matsubara frequencies, $%
\hbar $ is the reduced Planck constant, $r^{\alpha }(\alpha=s, p)$ are the
reflection coefficients for the DNA film, where the superscripts $\alpha=s$
and $p$ correspond to the polarizations of transverse electric ($\mathbf{TE}$%
) and transverse magnetic ($\mathbf{TM}$) modes, respectively. The
subscripts $1$ and $2$ denote the reflection coefficients at the top and
bottom interfaces of DNA film, respectively.

The reflection coefficients for an electromagnetic wave incident from the
DNA film to a medium (with permittivity $\varepsilon_1$) is given as \cite%
{Kli:20}:
\begin{eqnarray}
r^{\mathrm{TM}} &=&\frac{\varepsilon _{1}(i\xi _{n})k_{3}(i\xi
_{n},k_\parallel)-\varepsilon _{\mathrm{D}}(i\xi _{n})k_{1}(i\xi
_{n},k_\parallel)}{\varepsilon _{1}(i\xi _{n}) k_{3}(i\xi
_{n},k_\parallel)+\varepsilon_{\mathrm{D}}(i\xi _{n})k_{1}(i\xi
_{n},k_\parallel)} \\
r^{\mathrm{TE}} &=&\frac{k_{3}(i\xi _{n},k_\parallel)-k_{1}(i\xi
_{n},k_\parallel)}{k_{3}(i\xi _{n},k_\parallel)+k_{1}(i\xi _{n},k_\parallel)}
\end{eqnarray}
where $k_{1}=\sqrt{k_{\parallel }^{2}+\varepsilon _{1}(i\xi_{n})\xi
_{n}^{2}/c^{2}}$ is the vertical wavevector in the medium 1. Here, the
medium 1 can be the air, water, silica or gold.

The reflection coefficients are strongly dependent on the permittivity at
different Matsubara frequencies. Here, the dielectric functions of used
materials are fitted by a model of the modified harmonic oscillator, which
is adopted from a recent literature\cite{Moa:21}:
\begin{equation}
\varepsilon (i\xi )=1+\underset{j}{\overset{}{\sum }}\frac{C_{j}}{1+(\xi
/\omega _{j})^{\beta_j}},
\end{equation}%
where $C_{j}$ corresponds to the oscillator strength for the $j$-th
resonance frequency $\omega _{j}$, $\beta_j$ is a power exponent.
In addition to the Kramers-Kronig relations, the influences of the
electronic dielectric constant, optical bandgap, density, and chemical
composition are taken into account in the Eq.(4), where the parameters for
the dry DNA, water and silica are shown in Table 1.
\begin{table}[ht]
\caption{The parameters for the used materials \protect\cite{Moa:21}.}
\label{table:insulating}
\centering 
\begin{tabular}{p{2.0cm}p{2cm}p{2cm}p{1.5cm}}
\hline\hline
DNA: & $C_j$ & $\omega_j$(eV) & $\beta_j$ \\[0.5ex] \hline
$j$=1 & 1.766 & 0.0056 & 1.03 \\[1ex]
$j$=2 & 1.431 & 12.95 & 1.67 \\[1ex] \hline
Water: &  &  &  \\[0.5ex] \hline
$j$=1 & 73.48 & 8.1$\times10^{-5}$ & 0.988 \\[1ex]
$j$=2 & 2.534 & 0.016 & 1.1 \\[1ex]
$j$=3 & 0.755 & 16.1 & 1.751 \\[1ex] \hline
Silica: &  &  &  \\[0.5ex] \hline
$j$=1 & 1.843 & 0.0725 & 1.678 \\[1ex]
$j$=2 & 1.105 & 15.33 & 1.71 \\[1ex] \hline
\end{tabular}
\end{table}

It is worth mentioning that the dielectric function of the dry DNA given by
the parameters in Table 1 matches the measured data of DNA over a wide range of
frequencies, from zero frequency to the far ultraviolet \cite%
{Moa:21, Wit:86,Ina:74,Wei:87,Pau:18}. Based on the Clausius-Mossotti equation, the
permittivity for a wet DNA (denoted by $\varepsilon _{\mathrm{D}}^{^{\prime
}}$) is given by the following form \cite{Bar:19,Hou:80}:
\begin{equation}
\frac{\varepsilon _{\mathrm{D}}^{^{\prime }}(i\xi _{n})-1}{\varepsilon _{%
\mathrm{D}}^{^{\prime }}(i\xi _{n})+2}=\Phi \frac{\varepsilon _{\text{w}%
}(i\xi _{n})-1}{\varepsilon _{\text{w}}(i\xi _{n})+2}+(1-\Phi )\frac{%
\varepsilon _{\mathrm{D}}(i\xi _{n})-1}{\varepsilon _{\mathrm{D}}(i\xi
_{n})+2},
\end{equation}
where $\varepsilon _{\text{w}}(i\xi _{n})$ is the permittivity of
water, and $\Phi $ is the volume fraction of water in the DNA film.

The dielectric function of gold is given by summing up
the Drude model and the modified harmonic oscillator, which is written as:
\cite{Moa:21}
\begin{equation}
\varepsilon (i\xi )=1+\frac{C_{1}}{1+(\xi /\omega _{1})^{\beta_1}}+\frac{%
\omega_{p}^2}{\xi^2+\gamma \xi},
\end{equation}
where the parameters $C_1$=6.5, $\omega _{1}$=5.9 (eV), $\beta_1$=1.42, $%
\omega_{p}$=9.1 (eV), and $\gamma$=0.06 (eV). We find that the Casimir
calculations based on the gold permittivity in the Eq.(6) are the same as
those given by the generalized Drude-Lorentz model\cite{Ge:20a}.

\section{Results and discussions}

Figure 1(a) shows the permittivity of the applied materials evaluated in the
imaginary frequency. The results show that the permittivity of a dry DNA is
larger than those of the water and silica for the Matsubara term $n>0$. The
permittivity of water is the smallest over a wide range of frequencies. Figure 1(b) shows the permittivity of DNA under different water volume fractions. As expected, the permittivity of a wet DNA decreases by increasing the magnitude of $\Phi$, due to the elevated contribution from the low-refractive-index water.

To predict the sign of Casimir pressure, the permittivity at $n=0$ is significant since it plays a dominant role at a
large thickness (or separation) as reported in \cite{Bar:19, Est:16}. The
static permittivity for the silica, DNA and water are about 3.9, 4.2 and 81,
respectively. However, the dielectric function of a wet DNA film at $n=0$ shows a different trend, compared with the high-frequency one. The static permittivity $\varepsilon _{\mathrm{D}}^{^{\prime}}$ increases from 4.2 to 11.9, with increasing $\Phi$ from 0 to 0.6.

\begin{figure}[tbp]
\centerline{\includegraphics[width=8.2cm]{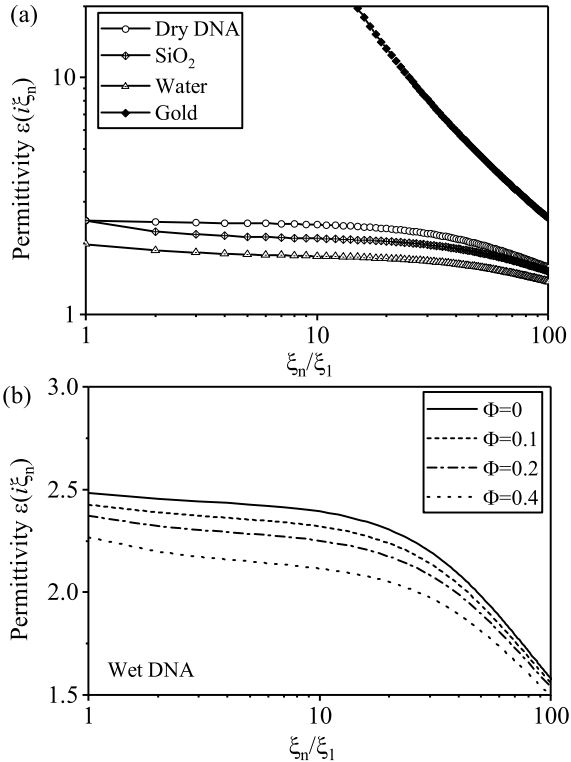}}
\caption{(a)The permittivity of used materials evaluated in
imaginary frequency. (b)The permittivity of wet DNA containing different
volume fractions of water.}
\label{Fig:fig1}
\end{figure}

\subsection{The Casimir pressures for suspended DNA films}

We first consider the Casimir force of a DNA film suspended in a
homogeneous background medium. The Casimir force would be
attractive as reported for suspended peptide films
\cite{Kli:20}. The absolute Casimir pressure versus the thickness of
suspended DNA film is shown in Fig. 2(a), where the solid and dash lines
represent the background media to be the air and water, respectively. It
is found that the magnitude of Casimir pressure decreases monotonously by
increasing the DNA thickness. The Casimir pressure for the air is larger
than the case of the water at a small thickness, while it is smaller at a
larger thickness.

The sign and magnitude of the Casimir pressure are dependent on the
dielectric responses of materials. Considering the DNA film is surrounded by
medium 1 and medium 2, the Casimir pressure would be proportional to the permittivity contrasts of the media \cite{Dzy:61}
\begin{equation}
P_{c}\propto \left( \frac{\varepsilon _{\mathrm{1}}(i\xi )-\varepsilon _{%
\mathrm{D}}(i\xi )}{\varepsilon _{\mathrm{1}}(i\xi )+\varepsilon _{\mathrm{D}%
}(i\xi )}\right) \left( \frac{\varepsilon _{\mathrm{2}}(i\xi )-\varepsilon _{%
\mathrm{D}}(i\xi )}{\varepsilon _{\mathrm{2}}(i\xi )+\varepsilon _{\mathrm{D}%
}(i\xi )}\right) ,
\end{equation}%
where $\varepsilon _{\mathrm{1}}$ and $\varepsilon _{\mathrm{2}}$ is the
permittivity of the medium 1 and medium 2, respectively. We have $\varepsilon _{\mathrm{1}}(i\xi )=\varepsilon _{\mathrm{2}}(i\xi )=1,\varepsilon _{\mathrm{1}}(i\xi
)=\varepsilon _{\mathrm{2}}(i\xi )=\varepsilon _{\mathrm{w}}(i\xi )$ when
the DNA film is suspended in the air and water, respectively. The
permittivity contrasts between DNA film and the air are larger than those of water
for $n>$0. It is known that the high
frequency components are dominant for the calculation of Casimir force at a small separation\cite{Yan:10}. As a result, the Casimir pressure for the air is larger than the that of the water at a small DNA thickness. By contrast, the dielectric contrast between
DNA film and the water is much larger than that of the air at $n$=0, which
is the leading term for a large DNA-film thickness. Thus, there is no
surprise that the Casimir pressure in water environment is larger than the
configuration of the air for a large thickness (e.g., $a>$500 nm).

\begin{figure}[tbp]
\centerline{\includegraphics[width=8.0cm]{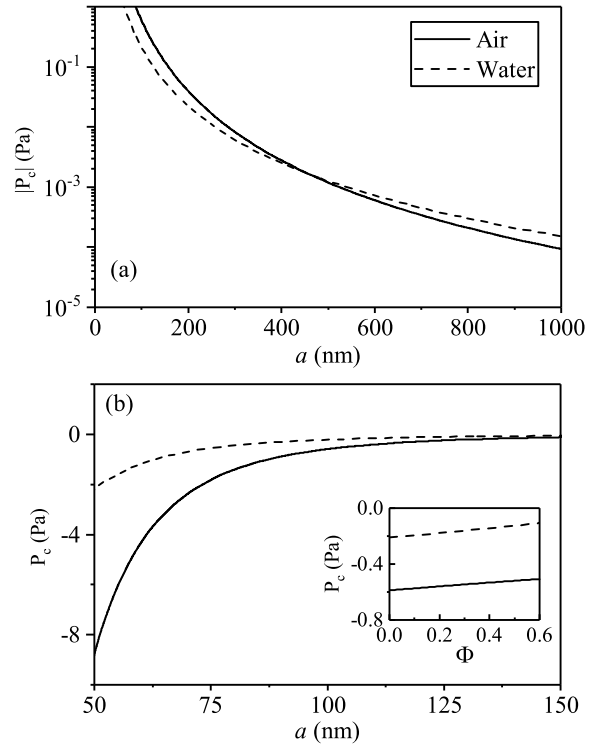}}
\caption{(a)The magnitude of Casimir pressure versus the
thickness of a dry DNA film suspended in the background of air
and water. (b) The Casimir pressure for thinner DNA films in
the linear scale. The inset shows the Casimir pressure as a function of
water volume fraction, where the DNA-film thickness $a$=100 nm is fixed. The
temperature is 300 K. }
\label{Fig:fig2}
\end{figure}

On the other hand, it would be interesting to consider the Casimir pressure
across a wet DNA film. As an
example, we set the thickness $a$=100 nm, and the Casimir force versus the
volume fraction of water in DNA film is shown in the inset of Fig. 2(b). The
results show that the magnitude increases with decreasing the value of $\Phi$%
. At the limit $\Phi$=0, the attractive Casimir pressures at the air and
water environment are about 0.2 and 0.6 Pa, respectively. The magnitude of
Casimir pressure is hundreds of times larger than the gravity of the DNA
film (about 1.7 mPa for $a$=100 nm), manifesting the important role of the
fluctuation-induced force. It can be seen that the magnitude of Casimir
pressure can be enlarged over 10 times, when the thickness $a$ decreases
further from 100 nm to 50 nm. Note that the condensation of DNA will decline
its thickness and the volume fraction of water (i.e., squeezing the water
out of the DNA film). Hence, it can be concluded that DNA films trends to
condensation for suspended configurations, due to the attractive Casimir
force.

\begin{figure}[tbp]
\centerline{\includegraphics[width=8.0cm]{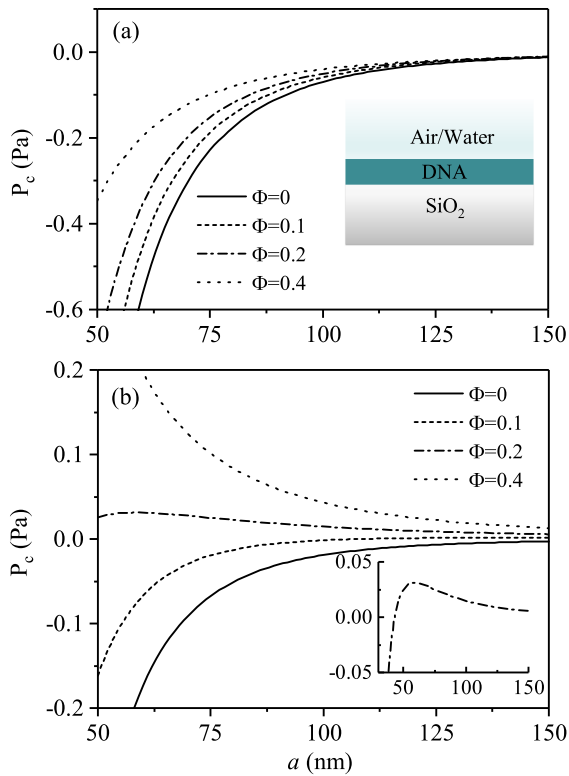}}
\caption{The Casimir pressure versus the thickness of DNA films
under different volume fractions $\Phi$. The substrate consists of
semi-infinite silica. (a)The DNA film is exposed to the air. (b) The DNA
film is immersed in the water. The inset in (b) shows the Casimir pressure
for $\Phi$=0.2 with a clearer plot scale. The positive (negative) sign
of the pressure corresponds to the repulsive (attractive) force.}
\label{Fig:fig3}
\end{figure}

\subsection{The Casimir pressures for a silica
substrate}

In many organic devices, the DNA film is generally deposited on a dielectric
substrate. The Casimir pressure for a DNA thin film placed on a silica
substrate is shown in Fig. 3(a), where the cladding background medium is the
air. The result shows that the Casimir pressure is negative, and its
magnitude increases by decreasing the DNA-film thickness. We note that
the magnitude of the Casimir pressure is about 0.07 Pa for the dry DNA film
at 100 nm, which declines considerably compared with the suspended
configuration (about 0.6 Pa). In addition, the magnitude of the Casimir
pressure for wet DNA film declines further with increasing volume fraction
$\Phi$. The pressure is only about 0.04 Pa with volume fraction $\Phi$%
=0.4. Nonetheless, the Casimir pressure for a wet DNA deposited on a silica
substrate is still much larger than the gravity of the DNA film. Overall, a
thin DNA film and a low water volume fraction are preferred for stability
of the DNA film.

As the DNA film is immersed in the water, some complicated or even reverse
conclusions are obtained, in comparison with the air configurations. The Casimir pressure as a function of the thickness $a$ is shown in Fig. 3(b). The results
show that the Casimir pressure is long-range negative at low volume fractions 0 and 0.1, and its magnitude decreases rapidly with increasing the DNA-film thickness. These properties suggest that a thin DNA film is favored for stability due to the attractive Casimir force, similar to the case of air in Fig. 3(a). However, the Casimir
pressure for a large $\Phi$ shows different features. For $\Phi$%
=0.4, the Casimir pressure is long-range positive, which
means that a thin thickness is harmful to the stability of DNA films. For an intermediate value $\Phi$=0.2, the Casimir pressure turns from negative to positive with
increasing the thickness $a$, as shown in the inset of Fig. 3(b). Then, a maximum peak
for the Casimir repulsion can be found near 60 nm, which contributes
negatively to the stability. The Casimir pressure would decrease with
increasing the thickness $a$ further.

The unusual behavior of Casimir pressure at the water background can be
interpreted by the competition between the attractive and repulsive Casimir
components. For a small $\Phi$, the permittivity of the wet DNA is larger
than those of silica and water over a wide range of frequencies ($n>0$). The
dielectric permittivity of the DNA and silica are very close at zero
frequency, resulting in a negligible contribution from the term $n=0$.
Therefore, the Casimir pressure is attractive according to the Eq. (7), and its magnitude increases rapidly with decreasing the DNA-film thickness, as demonstrated with $\Phi$=0
and 0.1 in Fig 3(b). For a large $\Phi$=0.4, the permittivity of the wet DNA is
smaller(larger) than that of silica(water) for $n>0$, resulting in repulsive Casimir force. At static frequency with $n$%
=0, the permittivity of the wet DNA is larger (smaller) than that of silica
(water), which also leads to repulsive Casimir force. Hence, the Casimir
pressure would be long-range repulsive for a large $\Phi$. For an
intermediate $\Phi$=0.2, the permittivity of the wet DNA is still larger
than that of silica(water) for $n>0$, resulting in attractive Casimir force at a small thickness $a$. However, the
contribution for $n$=0 is still positive, resulting in repulsive
Casimir force at a large value of $a$. Due to the competition between the
attractive and repulsive Casimir components, the peak for Casimir repulsion
is expected at an intermediate thickness, as shown in the inset of Fig. 3(b).

\begin{figure}[tbp]
\centerline{\includegraphics[width=8.2cm]{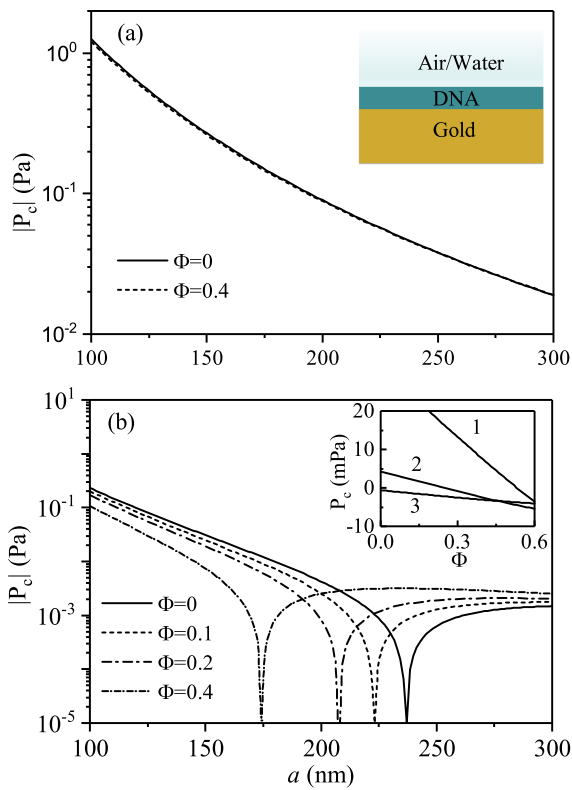}}
\caption{The magnitude of the Casimir pressure versus the
thickness of DNA films with a gold substrate. (a)The DNA film is exposed in
the air for $\Phi$=0 (solid) and 0.4 (dash). (b) The DNA film is immersed in
the water. The inset shows the corresponding Casimir pressure as a function
of $\Phi$. The labels 1, 2 and 3 represent the thickness of DNA film are 150
nm, 200 nm and 250 nm, respectively.}
\label{Fig:fig4}
\end{figure}

\subsection{The Casimir pressures for a metallic substrate}

Now we consider the case of DNA films deposited on the metallic substrate.
The magnitudes of the Casimir pressure as a function of thickness $a$ are
shown in Fig. 4(a), where the cladding medium is the air. According to Eq.(7), we can
predict that the Casimir pressure is long-range repulsive because $\varepsilon_{%
\mathrm{air}}<\varepsilon_\mathrm{D}'<\varepsilon_{\mathrm{Au}}$ is satisfied
for $n\geq0$ (see, e.g., Ref.\cite{Mun:09}). The magnitude
of Casimir pressure decreases monotonously with increasing the DNA-film
thickness. As a result, the dispersion forces make the DNA film less stable
for thin thickness. Note that the discrepancy of Casimir pressures acting on
the DNA film is small between volume fractions $\Phi$=0 and 0.4.

The Casimir pressure acting on the DNA film immersed in the water exhibit
different characteristic shown in Fig. 4(b). The Casimir pressure is
repulsive at a thin thickness, while it becomes attractive for a large thickness.
At a specific thickness, a stable Casimir equilibrium, i.e., the pressure equals to
zero, is found. The critical thickness for the Casimir equilibrium can be
modulated by the magnitude of $\Phi$. As the $\Phi$ increases from 0 to 0.4, the critical thickness decreases correspondingly from about 237 to 174 nm. The interesting Casimir equilibrium at the water background can be understood by the contrast of permittivity at the Eq. (7). The repulsive relation $\varepsilon_{%
\mathrm{w}}<\varepsilon_\mathrm{D}'<\varepsilon_{\mathrm{Au}}$ is satisfied for $n>0$, and the permittivity contrast between the water and wet DNA decreases with increasing the $\Phi$, resulting in a smaller Casimir repulsion.  On the other side, the attractive Casimir interaction at a large thickness $a$ is attributed to the relation $\varepsilon_{\mathrm{w}}>\varepsilon_\mathrm{D}'$ at the leading term $n=0$.

The inset in Fig. 4(b) shows the Casimir pressure changed by the volume
fraction of water with a fixed DNA-film thickness. We find that switching the sign of
the Casimir pressure from positive to negative is achieved by increasing
the volume fraction $\Phi$ for thickness 150 nm and 200 nm. For thickness 250
nm, the Casimir pressure is negative and its magnitude increases by
increasing the volume fraction $\Phi$. Hence, the DNA film deposited on the metallic substrate tends to be de-condensation in the water environment, according to the properties of its dispersion force.

\section{Conclusions}

In summary, the Casimir pressure of a DNA film is calculated in several
configurations based on the Lifshitz theory. The Casimir pressure is attractive when a DNA film is suspended in the air or water, and its magnitude increases with
decreasing the thickness of DNA film or/and the water volume fraction. Hence, the
suspended DNA film trends to condensation due to the Casimir force. The Casimir pressure is hundreds of times larger than the gravity of the DNA film for a moderate thickness (e.g., 100 nm), manifesting the important role of the fluctuation-induced interactions. For DNA films deposited on the silica substrate, the Casimir pressure is attractive for the air background. Also, a thin DNA film and a low water fraction are favored for
the stability. Instead, the Casimir pressure shows rich features in a water background.
The Casimir pressure can be changed from attractive to repulsive by increasing the DNA-film thickness and the water fraction. At the end, the Casimir
force of a DNA film deposited on a metallic substrate is explored. The Casimir pressure is dominant by the repulsive interactions at a small DNA-film thickness for both the air and water environment. For the setup immersed in a water environment, the Casimir pressure turns out to be attractive at a large DNA-film thickness, and a stable Casimir
equilibrium can be found at a specific thickness. Our finding provides a theoretical guide for the adhesion stability, condensation, and de-condensation of DNA films, resulting from the fluctuation-induced dispersion forces.

\begin{acknowledgments}
This work is supported by the National Natural Science Foundation of China
(Grant No. 11804288 and No. 61974127), and Shanghai
Pujiang Program (21PJ1411400). The research of
L.X. Ge is further supported by Nanhu Scholars Program for Young Scholars of
XYNU.
\end{acknowledgments}


\end{document}